\documentclass[conference]{IEEEtran}
\IEEEoverridecommandlockouts
\usepackage{cite}
\usepackage{amsmath,amssymb,amsfonts}
\usepackage{graphicx}
\usepackage{textcomp}
\usepackage{xcolor}
\usepackage{subfig}
\usepackage{booktabs}
\usepackage[ruled, vlined, linesnumbered]{algorithm2e}
\usepackage{cite}
\usepackage{longtable}
\usepackage{adjustbox}

\usepackage{color}

\def\BibTeX{{\rm B\kern-.05em{\sc i\kern-.025em b}\kern-.08em
    T\kern-.1667em\lower.7ex\hbox{E}\kern-.125emX}}

\begin{document}

\title{Multi-Agent RL-Based Industrial AIGC Service Offloading over Wireless Edge Networks}

\author{
    \IEEEauthorblockN{\large
    Siyuan Li$^{12}$\IEEEauthorrefmark{1}, 
    Xi Lin$^{12}$\IEEEauthorrefmark{1}\IEEEauthorrefmark{2}, 
    Hansong Xu$^{12}$\IEEEauthorrefmark{2},
    Kun Hua$^3$,
    Xiaomin Jin$^3$,
    Gaolei Li$^{12}$, 
    and Jianhua Li$^{12}$\\}
    \IEEEauthorblockA{$^{1}$\textit{School of Electronic Information and Electrical Engineering, Shanghai Jiao Tong University, Shanghai, China} \\}
    \IEEEauthorblockA{$^{2}$\textit{Shanghai Key Laboratory of Integrated Administration Technologies for Information Security, Shanghai, China} \\}
    \IEEEauthorblockA{$^{3}$\textit{Department of Electrical Engineering, California Polytechnic State University, San Luis Obispo, CA, USA} \\}
    \thanks{\IEEEauthorrefmark{1} These authors contributed equally.}
    \thanks{\IEEEauthorrefmark{2} Corresponding authors: Xi Lin (linxi234@sjtu.edu.cn) and Hansong Xu (hansongxu@sjtu.edu.cn).}
}

\maketitle

\begin{abstract}
Currently, the generative model has garnered considerable attention due to its application in addressing the challenge of scarcity of abnormal samples in the industrial Internet of Things (IoT).
However, challenges persist regarding the edge deployment of generative models and the optimization of joint edge AI-generated content (AIGC) tasks.
In this paper, we focus on the edge optimization of AIGC task execution and propose GMEL, a generative model-driven industrial AIGC collaborative edge learning framework.
This framework aims to facilitate efficient few-shot learning by leveraging realistic sample synthesis and edge-based optimization capabilities.
First, a multi-task AIGC computational offloading model is presented to ensure the efficient execution of heterogeneous AIGC tasks on edge servers.
Then, we propose an attention-enhanced multi-agent reinforcement learning (AMARL) algorithm aimed at refining offloading policies within the IoT system, thereby supporting generative model-driven edge learning.
Finally, our experimental results demonstrate the effectiveness of the proposed algorithm in optimizing the total system latency of the edge-based AIGC task completion.
\end{abstract}


\section{Introduction}
The data-driven discriminative AI models have been widely applied in various industrial Internet of Things (IoT) scenarios, including defect detection and network intrusion detection.
However, due to the limited number of abnormal samples, traditional discriminative models inevitably produce biased learning results, making few-shot learning a key challenge.
Conventional simulation-based methods for modeling detection processes in the industrial IoT have limitations in frequent updates to adapt to evolving real-world environments. 
Moreover, these approaches have yet to address the scarcity of anomaly samples in real-world scenarios effectively.
The utilization of generative models within the IoT has garnered significant attention \cite{wang2023unified}. 
Generative models have been a prevalent approach in addressing the challenges associated with few-shot learning for abnormal samples, especially the diffusion model, which utilizes a progressive diffusion process to generate high-resolution and realistic images \cite{tang2023wireless, stablediffusion:rombach2022high}.

Leveraging generative models into edge networks, focusing on personalized AI-generated content (AIGC) services and safeguarding user privacy, has gained significant attention \cite{xu2023unleashing, ji2023mixup}. 
Cloud pre-trained models (such as GPT-4) can be initially trained in cloud data centers, and then users can access cloud services through the core network, but this may result in high latency \cite{wang2023guiding}. 
Therefore, deploying interactive AI generation services in mobile edge networks is considered a more practical option.
Specifically, users can access low-latency, interactive AIGC services through AIGC-enhanced mobile networks, that is, pre-trained models are downloaded to mobile devices for inference \cite{zhang2022toward, xu2023unleashing}.
In addition, edge servers support localized service requests, allowing pre-trained models to fine-tune at the network edge.
In addition, AIGC users do not need to send requests to cloud servers within the core network, and edge-deployed AIGC models help maintain user privacy and security.
\begin{figure}[!t]
    \centering
    \includegraphics[width=\linewidth]{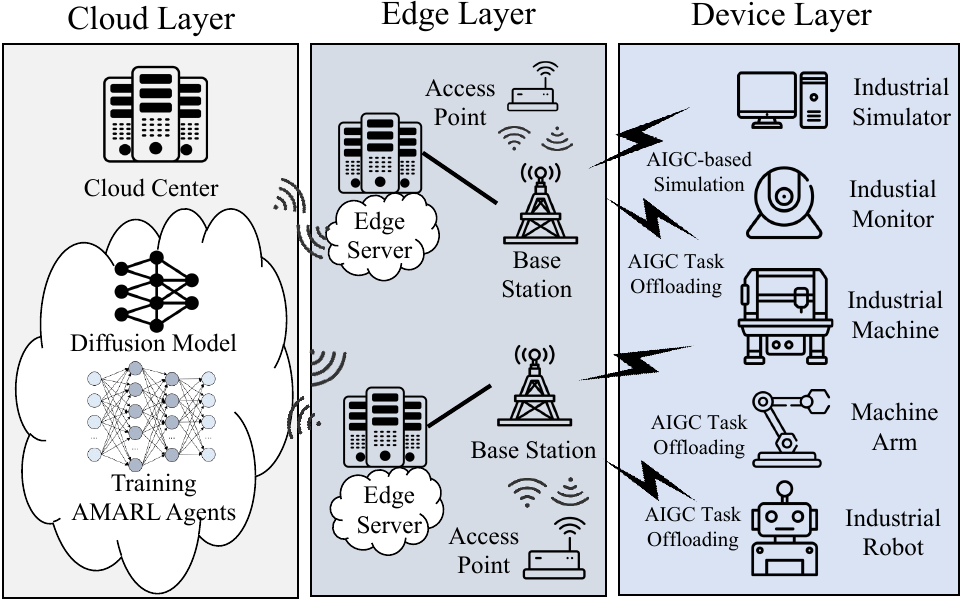}
    \caption{Overview of the proposed generative model-driven industrial AIGC collaborative edge learning framework.}
    \label{fig1}
\end{figure}

The integration of AIGC models into edge-based IoT necessitates the utilization of industrial simulators for industrial edge devices (IED) to enhance generative capabilities. 
These simulators leverage historical data under specific task conditions, thereby improving fidelity \cite{xu2023digital, xu2023generative}. 
Given the diverse computational resource requirements inherent in AIGC tasks, particularly on edge servers (ES), the optimization of the offloading process from IEDs to ESs becomes paramount for effective resource utilization \cite{lin2021stochastic}, as illustrated in Figure \ref{fig1}. 
Therefore, it is crucial to design algorithms that optimize the offloading process, aiming to minimize the system latency.

Recent research has explored multi-agent reinforcement learning (MARL) reinforcement learning for optimizing computing task offloading, especially in addressing the complex mixed cooperative-competitive interactions issues \cite{yao2023cooperative}.
Instead of each agent interacting with the environment, MARL considers multi-agent training at the same time, especially capturing the interaction features among IEDs.
Current research that applies MARL in mobile edge computing (MEC) still suffers from the problem that centralized critic networks can not address the vast joint state space and numerous useless information brought by other IEDs. 
The attention mechanism can effectively reduce the dimension of the joint state space and avoid invalid information from other IEDs, thereby enhancing resource utilization in communication and computing. 

To tackle these challenges, we propose an attention-enhanced multi-agent reinforcement learning (AMARL)-based AIGC tasks computing offloading algorithm, which is devised to offer efficient offloading policies to IEDs and optimize the allocation of resources to support AIGC-driven edge learning framework, as shown in Figure \ref{fig1}.
First, we propose GMEL, a novel AIGC-driven edge learning framework, with a diffusion-based generative model as a central component, as shown in Figure \ref{fig1}.
Then, within this framework, a multi-tasks AIGC tasks computational offloading model is correspondingly formulated to ensure the efficient execution of different AIGC tasks.
Finally, we propose a centralized training and decentralized execution MARL algorithm, while a multi-head attention mechanism is incorporated into the critic network during the centralized training period, to extract effective information in the complex industrial MEC environments.

The contributions of this work include:
\begin{itemize}
    \item 
    We design GMEL, an innovative AIGC-driven edge learning framework for industrial IoT, with a diffusion-based generative model as its core, which facilitates the edge optimization of providing real-time AIGC services.
    
    \item 
    We propose an AMARL-based algorithm aimed at resolving the computing offloading problem in industrial IoT, incorporating a multi-head attention mechanism to efficiently compress the joint observation space dimensions across all agents, ensuring network scalability.    
\end{itemize}

The rest of this paper is organized as follows: 
After discussing the related works in Section II, we discuss the system model and formulate the AIGC tasks computing offloading problem in Section III.
In Sections IV and V, the AMARL offloading algorithm and the experimental results are presented respectively. 
Finally, the conclusion is given in Section VI.

\section{Related Works}
Current AIGC models, especially advanced generative adversarial networks (GAN) and diffusion models present promising solutions for enhancing fault samples by utilizing training data. 
In the realm of IoT, particularly within few-shot learning scenarios, previous studies have explored the application of deep generative models. 
Addressing challenges associated with imbalanced data, Zhou \textit{et al.} introduce a distribution bias-aware collaborative GAN model for IoT scenarios \cite{zhou2022distribution}. 
Their approach involves a robust data augmentation framework designed to alleviate distribution bias between generated and original data, yielding significant improvements in multi-class classification distribution bias within IoT applications.
\begin{figure}[!t]
    \centering
    \includegraphics[width=\linewidth]{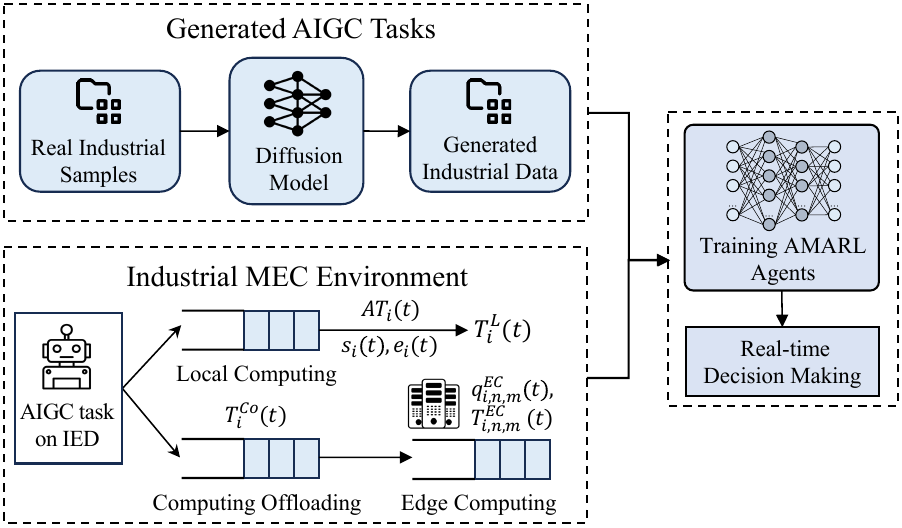}
    \caption{Structure of the AIGC-assisted industrial offloading decision and simulation.}
    \label{fig2}
\end{figure}

To enhance resource utilization in IoT, deep reinforcement learning (DRL) algorithms, notably MARL algorithms, are being developed to optimize the computing offloading process. 
Tang \textit{et al.} introduce a DRL-based distributed algorithm for decentralized task offloading in MEC systems, addressing challenges with uncertain edge node loads \cite{tang2020deep}. 
In the realm of the Internet of Remote Things, MARL is applied to coordinate computing task scheduling, such as optimizing offloading strategies in a space-air-ground integrated network \cite{zhang2023multi}. 
Despite their significance for intelligent computing offloading solutions, practical implementations of these methods in industrial scenarios remain limited.

\section{AIGC Tasks Edge Computing in Industrial IoT}
In this section, as shown in Figure \ref{fig1}, the generative model-driven industrial AIGC edge learning framework, GMEL, is devised to support the computing and communication resource utilization optimization with adherence to task-specific requirements \cite{li2023digital}.
Under GMEL, ESs with communication and computing capabilities are significant for providing AIGC computing services, enabling IEDs to offload AIGC tasks for remote execution while meeting strict deadlines.

\subsection{Edge-based Industrial AIGC Tasks for Simulation}
For the AIGC-based industrial simulation process, we adopt a combination of online simulation and offline training.
As shown in Figure \ref{fig2}, the AI-based simulator utilizes the available resources of ES to complete the training of the AIGC model and system decision model through offline training.
In these online environments, industrial simulators collect data from IEDs, while AIGC models generate different industrial samples for system decision modules, thereby enhancing the training process of these devices \cite{du2024diffusion}.
This combination of online execution and offline training helps to improve the industrial security and efficiency of the system \c.

A random AIGC task arrival model is employed in our model. 
Each IED experiences task arrivals according to a Poisson distribution at the start of discrete time intervals.
In the system model, we denote the set of ESs, wireless channels and IEDs as $\mathcal{M}=\{1, \ldots, m, \ldots, M\}, \mathcal{N}=\{1, \ldots, n, \ldots, N\}, \mathcal{I}=\{1, \ldots, i, \ldots, I\}$, respectively. 
We assume that an episode comprises $T$ consecutive time intervals. 
At the start of each time interval denoted as $t$, the generated AIGC task for IED $i$ is characterized by $AT_i(t) =\left\{s_i(t), e_i(t), d_i(t)\right\}$, where $s_i(t)$ is the size of the task graphic and other data, $e_i(t)$ is the required GPU cycles per unit data, and $d_i(t)$ signifies the maximum tolerable time before the deadline.
If a task is not fully transmitted or executed before its deadline, it will be abandoned.
Tasks within different time intervals may exhibit heterogeneity, meaning they possess different data sizes, resource requirements, and deadlines. 

To support services such as the execution of AIGC tasks and simulation of industrial processes, IED $i$ is endowed with GPU frequency $f_i^G$.
When a computing task is generated, IED needs to determine whether to execute locally or perform task offloading based on the current state of the IoT system.
We define $ \alpha_i(t)$ to indicate whether the task generated on the IED $i$ at time $t$ is being offloaded.
When the task needs to be offloaded to the ES, it is also necessary to determine the specific ES $m$ to be offloaded.
We define $\beta_{i, m}(t)$ to indicate whether $AT_i$ is to be offloaded to the ES $m$.

\subsection{Computing Latency in IoT}
First, for the local computing tasks, it is assumed that each IED encompasses a task queue.
The local computation latency $T_i^{L}(t)$ of task $AT_i(t)$ is mainly influenced by the computing resources of IEDs, the required GPU cycles per unit of data $e_i(t)$ and the local computation queuing latency.
In this model, AIGC tasks are considered to comprise a substantial portion of GPU-based simulations.
According to \cite{qiu2020distributed}, if the execution of tasks satisfies first-in-first-out, the task $AT_i$ will be completed in $\tau_{i}^{L}$-th time interval:
\begin{equation}
    \tau_{i}^{L}(t) = \max_{\hat{t_1}}\tau_i^{L}(\hat{t_1}) 
    +  \lceil \frac{s_i(t) e_i(t)}{\Delta t \cdot f_i^G}\rceil,
\end{equation}
where $\hat{t_1} \in\{1, \ldots, t-1\}$, $\Delta t$ denotes the time of a single time interval and $\lceil\cdot\rceil$ is the rounding up operation.
Therefore, for the task $AT_i$, the local computation latency of IED $i$ $T_i^{L}(t)$ can be calculated as $(\tau_i^{L}(t) - t)\Delta t$.

As shown in Figure \ref{fig2}, for tasks offloaded to the ESs, ESs execute the received AIGC tasks using their computing resources.
The computing latency primarily depends on the data size of the offloaded task and the allocated resources available to the IED.
In our computing model, each ES maintains $i$ ES computation queues.
As shown in Figure \ref{fig2}, similar to the local section, we consider GPU-based simulation.
The length of the ES computation queue is denoted as $q_{i,n,m}^{EC}(t)$, and the computing capability of ES $m$ is denoted as $F_m^G$.

The ES computation queue $\mathcal{V}_m^{EC}(t)$ is defined as valid if tasks reach the ES $m$ or the task in the queue was not completely processed in the time interval $t-1$.
Let $\xi_{i,n,m}^{EC}(t)$ denote the amount of graphic data discarded by the valid ES computation queue at the ending of time interval $t$. 
The length of the valid ES computation queue can be computed as:
\begin{equation}
    q_{i,n,m}^{EC}(t) = 
    q_{i,n,m}^{EC}(t-1)
    - \frac{F^G_m \Delta t}{|\mathcal{V}_m^{EC}(t)| e_i(t)}
    - \xi_{i,n,m}^{EC}(t) + s_{i,m}(t),
\end{equation}
which means the queue length equals the queue length of the time interval $t-1$ minus the number of completed and discarded tasks and the number of new arrivals.

Based on the above discussion, the edge computing latency $T_{i, n, m}^{EC}(t)$ in processing the AIGC tasks $AT_i(t)$ of IED $i$ in ES $m$ can be calculated using the time interval between computing starting time $\tau_{i,m,0}^{EC}(t)$ and completion (or discarding) time $\tau_{i,m}^{EC}(t)$.

\subsection{Communication Latency in IoT}
For the situation that the ED $i$ offloads the task $AT_i(t)$, it will first send the task to the selected ES $m$.
The transmission time of the returned data is neglected.
OFDMA is considered as the multiple access technology to partition the system bandwidth $B$ into equivalent sub-bands allocated to each IED.

In addition, besides the degradation of path loss, the dynamically changing channel gain is also considered in our model, to better adapt to the actual industrial environment.
According to \cite{gao2022large}, the normalized channel gain $\bar{g}_{i,m}(t)$ for IED $i$ can be calculated using all the channel gain vectors of IEDs.
Similar to the edge computing situation, the transmission of the offloaded tasks in a channel can be modeled as a communication queue.
We denote the length of the communication queue in channel $n$ from IED $i$ to ES $m$ as $q_{i,n,m}^{Co}(t)$, and define the set of the active valid communication queue of IED $i$ at time interval $t$ as $\mathcal{V}_\mathcal{N}^{Co}(t)$.

In our network mode, it is assumed that each IED's valid communication queue will be allocated with average bandwidth to guarantee each IED uses the transmission resource equally. 
In addition, the transmission power of IED $i$ is represented by $P_i$.
Therefore, using the number of valid wireless communication queues, the channel $n$ uplink transmission rate $R_{i,n,m}(t)$ between IED $i$ and ES $m$ can be computed.
Similar to the edge computing situation, the length of the valid wireless communication queue $q_{i,n,m}^{Co}(t)$ can also be computed using the length in $t-1$ interval $q_{i,n,m}^{Co}(t-1)$, $s_i(t)$, and $R_{i,n,m}(t)$.
Moreover, the communication latency can also be defined similarly.

\subsection{Problem Formulation}
It is necessary to coordinate the competitive and cooperative relationships between each IED to achieve global optimization. 
Otherwise, interference between IEDs may lead to resource waste, long latency, and incomplete tasks.
In this article, this is described as a coordination problem for multi-agent systems.
Specifically, this paper considers trying to find the optimal AIGC task computation offloading strategy under the constraints of available resources, including channel resources and maximum latency before the deadline, to minimize overall system latency, which can be formulated as:
\begin{equation}
    \operatorname{Min} \sum_{i \in \mathcal{I}} (1 - \alpha_i(t)) T_i^{L}(t) + \alpha_i(t)\left(T_i^{Co}(t) + T_{i,m}^{EC}(t)\right),
    \label{formulation} 
\end{equation}
\begin{align}
    \text{s.t.} \enspace & \sum_{i \in \mathcal{I}} B_{i, n}(t) \leq B (i \in \mathcal{V}_{\mathcal{N}}^{Co}(t)), \label{formulation_a} \vspace{1ex} \tag{3a} \\
     & T_i^{L}(t) \leq d_i(t),
    T_i^{Co}(t) + T_{i, m}^{EC}(t) \leq d_i(t). \label{formulation_b} \tag{3b}
\end{align}

\section{Attention-based MARL for AIGC Computing Tasks Offloading}
The problem defined by equation (\ref{formulation}) is hard to process by conventional methods, and we address it in this paper by leveraging an effective MARL-based solution. 
In this section, we elaborate on the multi-agent model of the state, action, and reward of IEDs, and introduce the proposed AMARL-based AIGC computing offloading algorithm.

\subsection{Formulating the Multi-agent Model}
\subsubsection{State}
The local observation of IED $i$ at time step $t$ is denoted as $\boldsymbol{o}_i(t)$, including the information of its generated AIGC task and the perceptible current network states.
IED $i$ can only observe the information of AIGC task $AT_i(t)$ generated on it, which includes the size of the task graphic and other data $s_i(t)$ the required GPU cycles per unit data $e_i(t)$, and the maximum tolerable time $d_i(t)$.
The perceptible current network states include the queuing latency of the local processing queue $T_i^{L}(t)$, the length of the valid wireless communication queue $q_{i,n,m}^c(t-1)$, the length of the valid ES computation queue  $q_{i,n,m}^{EC}(t-1)$, and the wireless channel state $\bar{g}_{i,m}(t)$ and $\omega_{i,m}$.

\subsubsection{Action}
The action of IED $i$ in the time interval $t$ is represented as $\boldsymbol{a}_i(t)$, which includes whether to offload the AIGC task and which ES is selected.
We introduce $\alpha_i(t)$ and $\beta_{i, m}(t)$ to indicate whether to offload the current task and whether ES is selected to offload the task, both of which are logic variables.
In conclusion, the action of the IED $i$ at time interval $t$ can be specified as $ \boldsymbol{a}_i(t) = \{\alpha_i(t), \beta_{i, m}(t) \}$.

\subsubsection{Reward}
In this work, we define the reward of the IED $i$ at the time interval $t$ as:
\begin{equation}
    r_i(t) = \sum_{t=1}^T \gamma^t (C - T_i(t)),
\end{equation}
where $\gamma \in[0,1]$ is the discount factor, $T_i(t)$ is the response latency of task $AT_i(t)$ and $C$ is a constant. 
Under the state $\boldsymbol{o}_i(t)$, IED $i$ adopts the offloading action $\boldsymbol{a}_i(t)$ to interact with the system and get the reward $r_i(t)$, and the goal of our following AMARL-based computing offloading algorithm is to maximize the $r_i(t)$ of all IEDs over a long period.

\subsection{AMARL-based computing offloading algorithm}
Our algorithm mainly tackles the challenges of partial observability in the MEC system and high-dimensional input space in applying RL to MEC by introducing an attention mechanism and several techniques. 
AMARL is trained using the centralized training and decentralized execution mechanism, primarily conducted in the cloud center of ample computing resources. 
During training, agents leverage collective observation data, but during decentralized execution, each IED $i$ solely relies on its partial observation $\boldsymbol{o}_i(t)$ to make offloading decision $\boldsymbol{a}_i(t)$.
Trained AMARL agents, deployed across IEDs, utilize their observations for decision-making during execution, with each agent using its partial observation as input for its actor network and obtaining reward $\boldsymbol{a}_i(t)$ from the system for iterative interaction and experience accumulation.
An attention mechanism is introduced to calculate the action-value function for each IED agent, encoding and processing states and actions of IEDs and others separately, ultimately training the AMARL using the policy gradient algorithm until convergence.

The replay buffer pool $E_i$ of agent $i$ consists of some stored experience tuples.
The IED agent $i$ operates its own operation $\boldsymbol{a}_i(t)$ based on its partial observation $\boldsymbol{o}_i(t)$ and the offloading strategy, and obtains the reward $r_i(t)$ from the system.
Let $h_{a}$ and $h_{c}$ indicate the hidden status of the actor network and the critic network respectively.
The value function of IED agent $i$ is denoted as $Q_i^{\pi}(O(t), A(t), h_c(t))$, where 
$O(t) = {\boldsymbol{o}_i(t), i \in \mathcal{I}}$, and $A(t) = {a_i(t), i \in \mathcal{I}}$, so the policy gradient update can be defined as:
\begin{equation}
    \begin{aligned}
    \nabla_{\theta_i} \mathcal{J}(\boldsymbol{\pi}) 
    = & \mathbb{E}_{E}
    [\nabla_{\theta_i} \pi_i (\boldsymbol{a}_i(t) 
    \mid \boldsymbol{o}_i(t), 
    h_a(t) ) \\
    & \cdot \nabla_{\boldsymbol{a}_i(t)} 
    Q_i^{\pi}(O(t), A(t), h_c(t))],
    \end{aligned}
\end{equation}
where $\boldsymbol{a}_i(t) = \pi_i (\boldsymbol{o}_i(t), h_a(t))$.

In a large-scale IED environment, the critic networks of agents face the problem of excessive state space caused by excessively invalid information from other IEDs.
To solve this problem, we introduce an attention mechanism to compress the dimension of the joint state space after synthesizing the states of all other IED agents.
Specifically, the action-value function of IED agent $i$ is $Q_i^\pi(O(t), A(t), h_c(t))$.
We can calculate the action-value function of the IED agent $i$, which other IED agents jointly influence:
\begin{equation}
    Q_i^{\pi} (O(t), A(t), h_c(t)) 
    = \varphi_i \left(\eta_i \left( \boldsymbol{o}_i(t), \boldsymbol{a}_i(t), h_c(t)\right), \omega_i\right),
\end{equation}
where $\varphi_i$ is the fully-connected layers, $\eta_i$ is the convolutional layers.
The impact of other IED agents on IED agent $i$ is achieved through $\omega_i$ \cite{iqbal2019actor}.

When training using the attentive critic network, all critics are updated simultaneously to minimize the collaborative regression loss function since the parameters are being shared:
\begin{equation}
    \mathcal{L}_Q(\pi)
    = \sum_{i=1}^{I} \mathbb{E}_{E} 
    (Q_i^{\pi} (O(t), A(t), h_c(t)) - y_i )^2,
\end{equation}
where
\begin{equation}
    \begin{aligned}
    y_i = \enspace
    & r_i(t) + \gamma \cdot E_{\pi_{\theta_i^{\prime}} } [ Q_i^{\pi^{\prime}} (O^{\prime}(t), A^{\prime}(t), h_c^{\prime}(t) ) \\
    & - \lambda^{\prime} \log (\pi_{\theta_i^{\prime}} (\boldsymbol{a}_i^{\prime}(t) \mid \boldsymbol{o}_i^{\prime}(t), h_a^{\prime}(t) ) ) ],
    \end{aligned}
\end{equation}
where $\pi^{\prime}$ indicates the parameters of the target critic network. 
$\theta_i^{\prime}$ denotes the parameters of the target actor network. 
It is worth noting that $Q_i^{\pi}$ estimates the action value for IED agent $i$ by accepting observations and actions for all IED agents. 
$\lambda^{\prime}$ is the constant that controls the trade-off between entropy and return maximization. 
\begin{figure}[!t]
    \centering
    \includegraphics[width=0.85\linewidth]{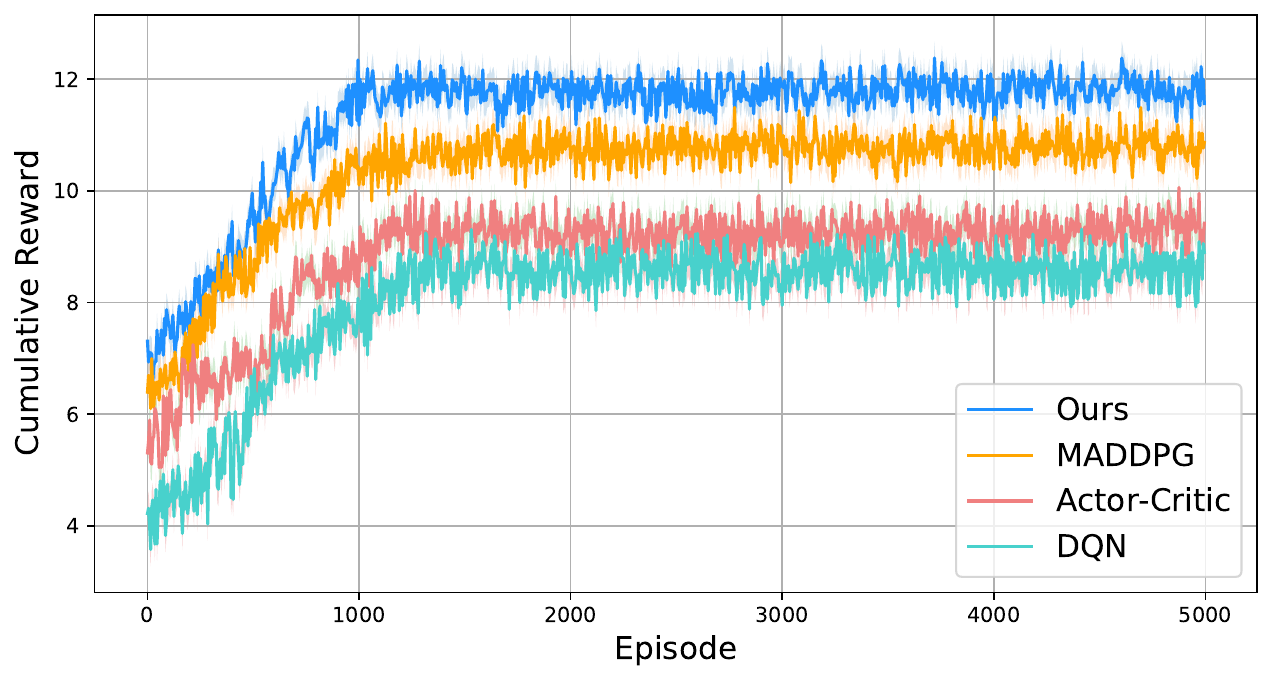}
    \caption{Convergence and cumulative rewards of different algorithms vs. episodes}
    \label{fig3}
\end{figure}
\begin{figure}[!t]
    \centering
    \includegraphics[width=0.85\linewidth]{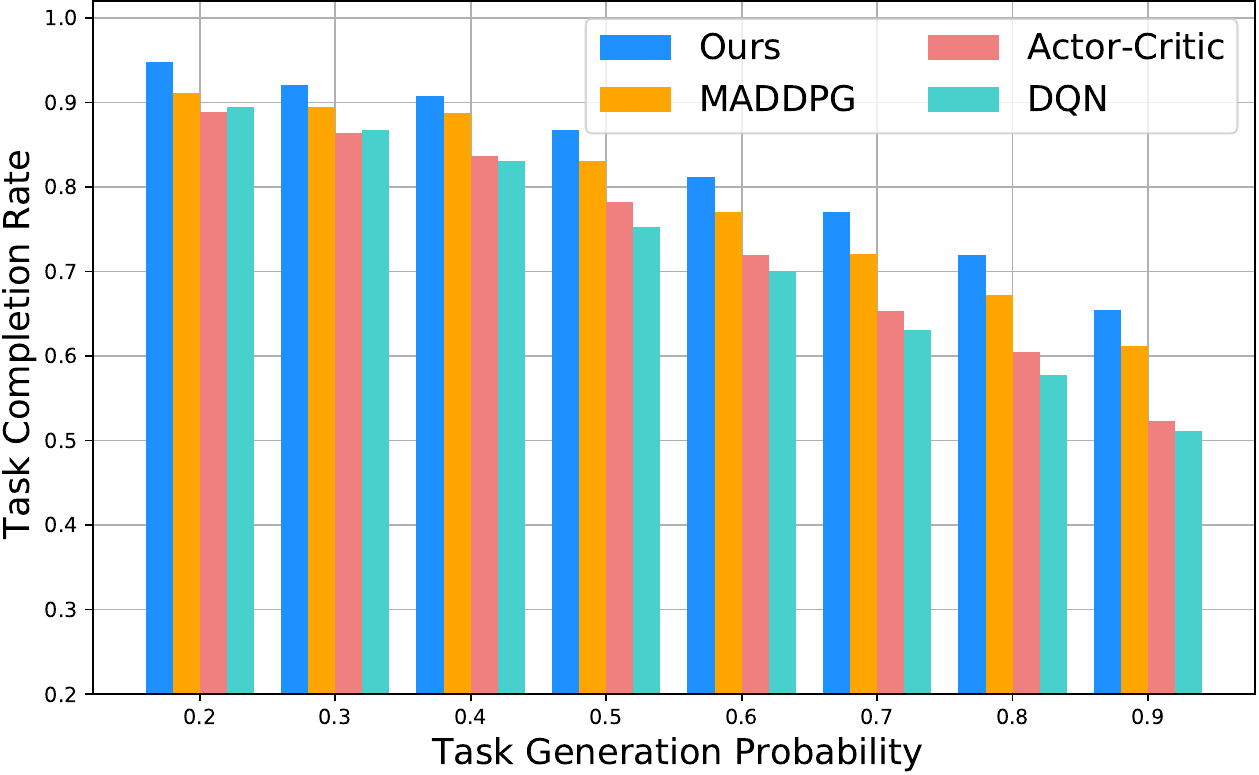}
    \caption{Task completion rate vs. task generation probability}
    \label{fig4}
\end{figure}
\begin{figure}[!t]
    \centering
    \includegraphics[width=0.85\linewidth]{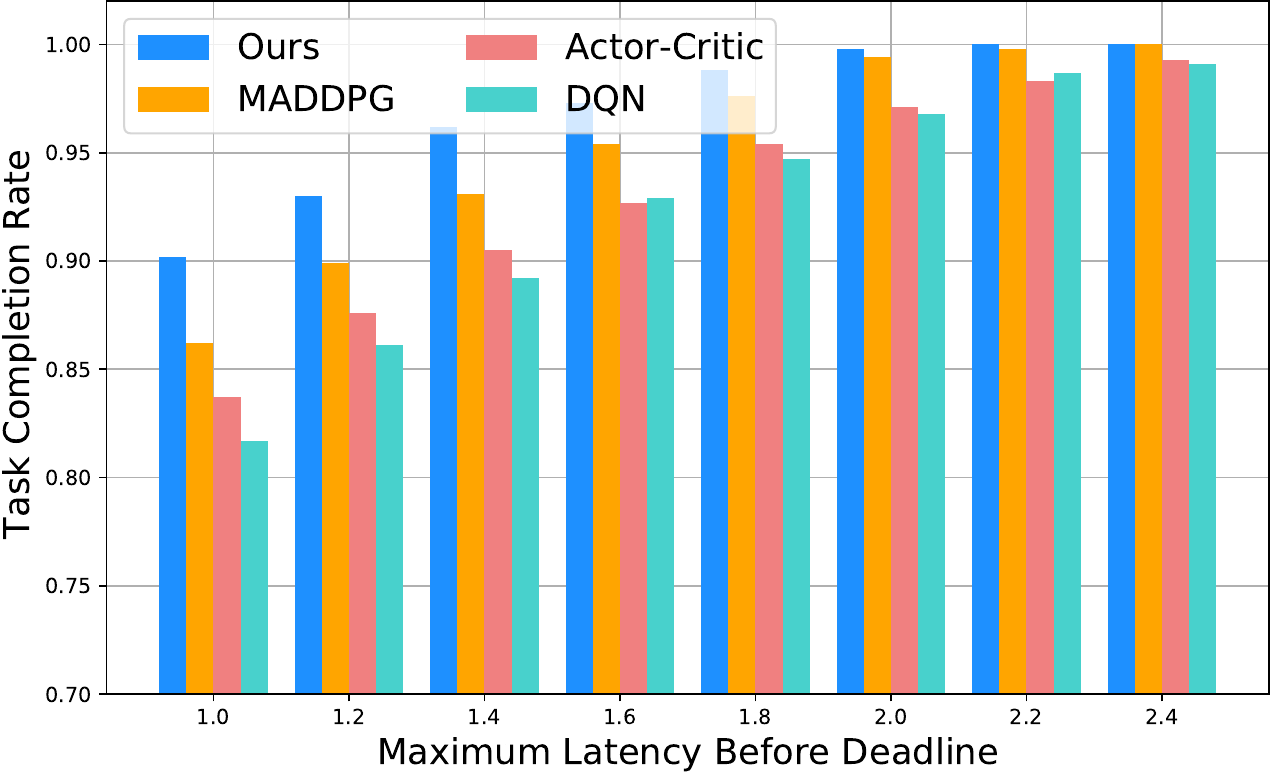}
    \caption{Task completion rate vs. maximum latency before deadline}
    \label{fig5}
\end{figure}
\begin{figure}[th]%
    \centering   
\subfloat[]{\includegraphics[width=0.485\linewidth]{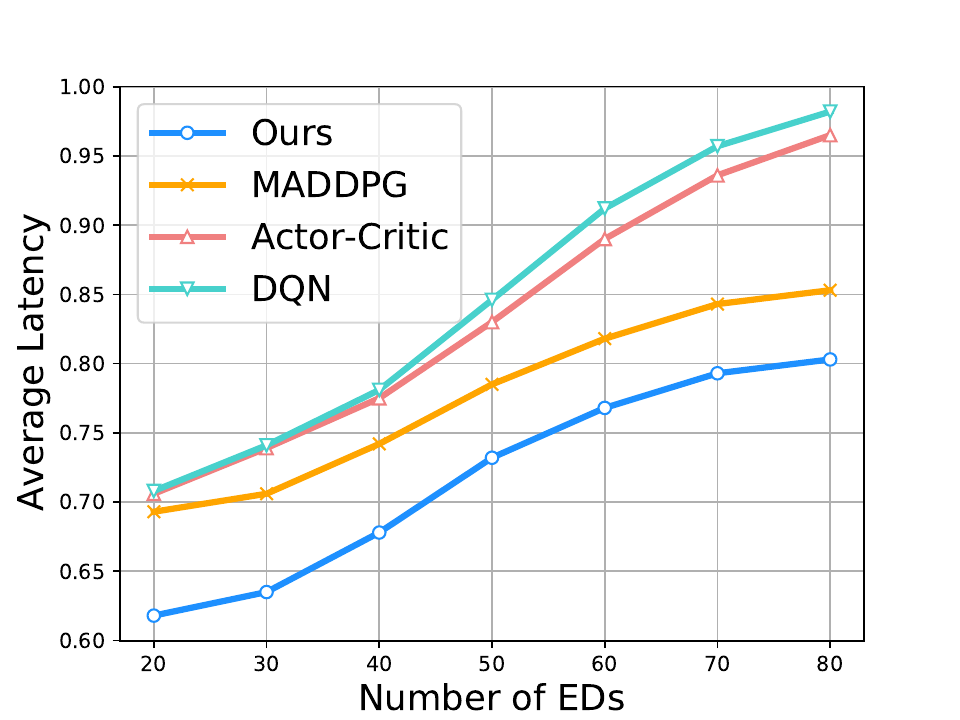}\label{fig6-a}}
    \hspace{0.01\linewidth}
\subfloat[]{\includegraphics[width=0.485\linewidth]{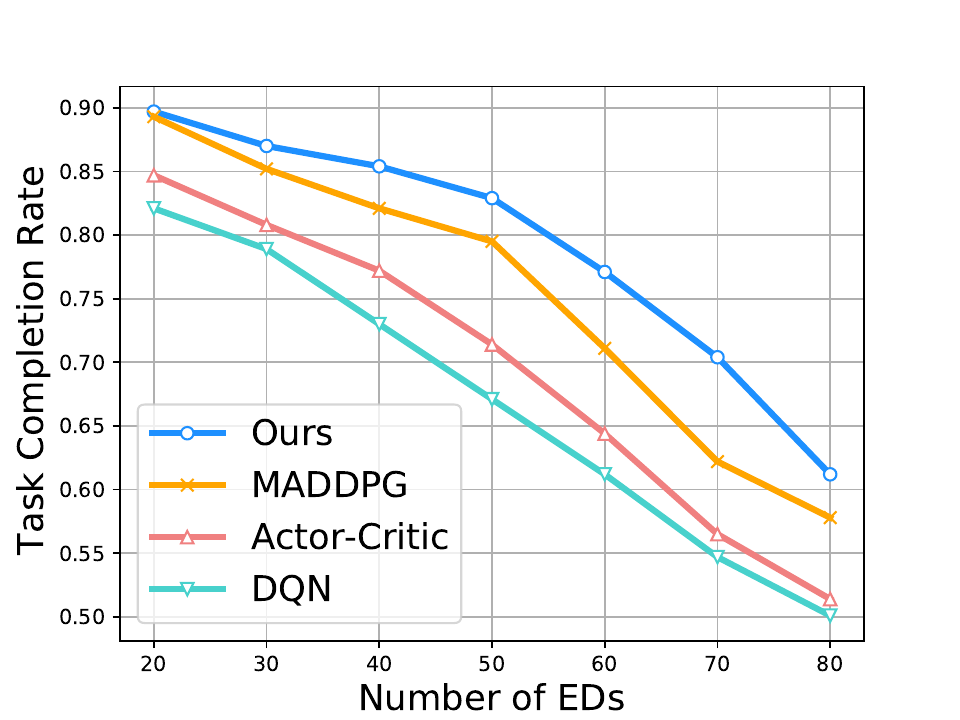}\label{fig6-b}}
\caption{Number of IEDs vs. (a) average Latency and (b) task completion rate}
\label{fig6}
\end{figure}

In addition, smoothing of target policies is applied, incorporating Gaussian distribution noise $\zeta = 
\mathcal{N}(0, \sigma)$ into the actions of all IED agents in the critic network update. 

The value function $Q_i^{\pi}(O(t), A(t))$ of the critic network is divided into two sections. 
In the dueling critic network, we add two sub-network structures, corresponding to the value-function section and the advantage function section. 
The output of the final critic network is obtained by linearly combining their output.

\section{Experiment}
In this section, we present the effectiveness of the proposed AMARL-based AIGC computing offloading algorithm. 
First, the experimental setups corresponding to the previous modeling are given in Section V-A.
Then, in Section V-B, the performance of the proposed algorithm with the existing methods under different system settings is evaluated.

\subsection{Experimental Setups}
First, our simulated industrial IoT environment parameters are introduced.
We consider a 100m $\times$ 100m industrial IoT system with 50 IEDs and 5 ESs by default. 
For each AIGC task generated by IED, the data size $s_i(t)$ is randomly sampled from 0.5 to 5 MB, the $e_i(t)$ are randomly sampled from 0.1 to 0.5 G cycles/MB, and the required deadline $d_i(t)$ is randomly sampled from 0.5 to 2.5 seconds.
In addition, the GPU computing capacity of ES $F_m^G$ varies from 10 to 20 GHz. 
The range of values for system channel bandwidth $B$ for AIGC task transmission is 10-20 MHz. 
The duration of the time interval is set to 0.1 s.
The additive white Gaussian noises at IEDs and ESs are randomly sampled from $N(0, 1)$, where $N$ denotes the normal distribution. 
For the AMARL model, the learning rate of the actor and critic is set to 0.001 and 0.005, respectively.
Moreover, we set the discount factor $\gamma$ as 0.99, the batch size to 64, and the dropout rate to 0.1. 

\subsection{Evaluation of AMARL-based Computing offloading Algorithm}
To validate the effectiveness of the AMARL model, we assessed the AMARL-based computation offloading method through 5000 episodes.
It is important to note that each iteration was configured with 100 time intervals, where each time interval signifies one interaction between each IED agent and the MEC environment.
Figure \ref{fig3} shows the convergence and the average reward of our algorithm compared to the existing DRL algorithms, MADDPG, Actor-Critic, and DQN.
Compared with these three benchmark offloading methods, especially the MADDPG-based algorithm, our proposed AMARL-based computing offloading algorithm exhibits relatively fast convergence and achieves higher rewards, which shows the attention mechanism effectively handles vast state space.

In Figure \ref{fig4}, we investigate the correlation between the probability of task generation and task completion rate as well as average system cost, to prove that our algorithm effectively reduces transmission latency and improves the completion rate of offloaded AIGC tasks.
It is worth noting that the computing offloading method based on the AMARL model has also outperformed the computation offloading method based on other DRL algorithms, which also reveals that introducing a multi-head attention mechanism is particularly effective. 
All algorithms at the start of the initial stage can reach a better performance; however, when the probability of task generation is high, our algorithm can always achieve a higher task completion rate than other algorithms. 
As shown in Figure \ref{fig4}, as the probability of task generation reaches 0.7, our algorithm outperforms other algorithms, particularly with a 6.93\% higher task completion rate compared to MADDPG.
Figure \ref{fig5} illustrates that the task completion rate increases with the maximum latency before the deadline gradually increases.
Our algorithm perpetually obtains a higher task completion rate than other benchmark algorithms, particularly while the task maximum latency before the deadline is short. 
For instance, when the maximum latency before the deadline is only 1.0 s, the AMARL outperforms the other benchmarks with the task completion rate increased by at least 4.64\%.

Moreover, the proposed algorithm outperforms the other algorithms when the number of IEDs increases, as demonstrated in Figure \ref{fig6}.
As the number of IEDs increases, the average latency of all the algorithms increases and the task completion rate decreases.
However, the AMARL consistently achieved the best results, which indicates that it can adapt to large-scale scenarios well.
What's more, the computation offloading algorithms based on DRL are typically better than some fixed policy methods, such as greedy allocation policy and average allocation policy. 
However, we believe that these methods are not solid baselines, so they are not included in the results.

\section{Conclusion}
In this paper, we proposed GMEL, a generative model-driven industrial AIGC collaborative edge learning framework to optimize computing and communication resource utilization and facilitate efficient industrial few-shot learning. 
Specifically, we propose a novel AMARL-based AIGC computing offloading algorithm, aimed at refining offloading policies within the industrial IoT system, thereby supporting generative model-driven edge learning. 
We improve the algorithm by incorporating a multi-head attention mechanism, to successfully reduce the impact caused by high-dimensional input of critic networks.
Experimental results demonstrate that our proposed algorithm consistently outperforms four other baselines.

\section*{Acknowledgment}
This work was supported by the National Natural Science Foundation of China under Grant 62202302, Grant 62202303, Grant U20B2048, and Grant 62302301.

\bibliographystyle{IEEEtran}
\bibliography{mine}

\end{document}